\documentclass[prd,showpacs]{revtex4}
\usepackage{amsmath,amssymb}
\usepackage{graphicx}
\usepackage{float}

\numberwithin{equation}{section}
\begin{document}
\title{Casimir effect with one large extra dimension}
\author{Andrea Erdas}
\email{aerdas@loyola.edu}
\affiliation{Department of Physics, Loyola University Maryland, 4501 North Charles Street,
Baltimore, Maryland 21210, USA}
\begin {abstract} 
In this work I study the Casimir effect of a massive complex scalar field in the presence of one large compactified extra dimension. I investigate the case of a scalar field confined between two parallel plates in the macroscopic three dimensions, and examine the cases of Dirichlet and mixed (Dirichlet-Neumann) boundary conditions on the plates. The case of Neumann boundary conditions is uninteresting, since it yields the same result as the case of Dirichlet boundary conditions. The scalar field also permeates a fourth compactified dimension of a size that could be comparable to the distance between the plates. This investigation is carried out using the $\zeta$-function regularization technique that allows me to obtain exact expressions for the Casimir energy and pressure. I discover that, when the compactified length of the extra dimension is similar to the plate distance, or slightly larger, the Casimir energy and pressure become significantly different than their standard three dimensional values, for either Dirichlet or mixed boundary conditions. Therefore, the Casimir effect of a quantum field that permeates a compactified fourth dimension, could be used as an effective tool to explore the existence of large compactified extra dimensions.
\end {abstract}
\maketitle
\section{Introduction}
\label{1}
Hendrik Casimir first recognized that an attractive force exists between two uncharged, conducting parallel plates in vacuum, entirely due to quantum effects \cite{Casimir:1948dh}. Sparnaay was first to attempt measuring this force \cite{Sparnaay:1958wg}, ten years after Casimir's original paper. Many increasingly accurate experiments followed throughout the decades \cite{Bordag:2001qi,Bordag:2009zz} that fully confirmed the Casimir effect, caused by quantum fluctuations of the electromagnetic field in a vacuum constrained by two conducting parallel plates. Fluctuations of quantum fields of different types also produce Casimir forces, that depend strongly on the boundary conditions at the plates of the field considered. For a scalar field, Dirichlet or Neumann boundary conditions cause attraction between the plates, while mixed (Dirichlet-Neumann) boundary conditions, cause a repulsive force \cite{Boyer:1974}. Casimir forces depend strongly also on the geometrical shape of the conducting plates, as discovered by Boyer \cite{Boyer:1968uf}, when he realized that spherical plates are subject to a repulsive force, caused by either the electromagnetic field, or a scalar field, satisfying Dirichlet boundary conditions at the spherical shell.

The existence of compactified extra dimensions has been proposed by Kaluza and Klein a century ago, as an attempt to unify gravity and electromagnetism \cite{Kaluza:1921tu,Klein:1926tv}. The idea has been revived twenty yers ago in a series of famous papers by Arkani-Hamed, Dimopoulos and Dvali \cite{Arkani-Hamed:1998jmv,Antoniadis:1998ig,Arkani-Hamed:1998sfv}, where large compact extra dimensions were proposed within a new framework, embedded in string theory, for solving the hierarchy problem. The extra dimensions size proposed by these authors should range between roughly a millimeter and a TeV$^{-1}$. One year later, Randall and Sundrum \cite{Randall:1999ee} found a different solution to the hierarchy problem involving a five-dimensional AdS spacetime with warped geometry.
All these scenarios modify gravity by producing deviations from Newton's law of gravitation at the submillimiter distance. Several experiments have been carried out in the last twenty years that test gravity at the submillimiter scale, see Refs. \cite{Adelberger:2009zz,Murata:2014nra} for a comprehensive list of experiments, placing deviations from Newton's gravity at or below $40\mu$m. Cosmological and astrophysical constraints also exist on the size and number of the compactified extra-dimensions, as well as collider constraints from LHC.

The Casimir effect can be a powerful experimental tool to search for the existence of a large compactified extra dimension, as long as an accurate theoretical investigation is carried out to understand exactly the implication of such extra dimension on the Casimir effect experiments in the lab. Several papers have been published on this subject, see Refs. \cite{Teo:2008ah,Teo:2009tm} for a comprehensive list, but the exact and detailed analysis that links compactification length, plate distance and mass of the field to deviation of Casimir energy and pressure from their well known three dimensional values, is missing. In this paper I will undertake that detailed analysis by investigating a massive scalar field constrained between two parallel plates in three dimensions and satisfying either Dirichlet or mixed boundary conditions on the plates. The scalar field also permeates a compactified fourth dimension sized at the submillimiter scale. I will use a well known regularization method introduced by Hawking \cite{Hawking:1976ja}, the zeta function technique, to regularize the Casimir energy \cite{Elizalde:1988rh, Elizalde:2007du}.

In Sec. \ref{2} of this work I describe the model of a massive scalar field permeating a compactified extra-dimension and obtain the exact zeta function in the two different cases of Dirichlet and mixed boundary conditions at the parallel plates. Sec. \ref{3} is dedicated to Dirichlet boundary conditions and I obtain exact and simple expressions for the Casimir energy in the two cases of a field with light mass and heavy mass respectively. In Sec. \ref{4} I investigate mixed boundary conditions and find exact and simple expressions of the Casimir energy for the two cases of light and heavy mass. In Sec. \ref{5} I use the Casimir energy obtained in the previous two sections to evaluate the Casimir pressure for Dirichlet and mixed boundary conditions in the two cases of light and heavy mass. A discussion of my results and my conclusions are in Sec. \ref{6}.
\section{The model and its zeta function}
\label{2}
In this paper I investigate a complex scalar field $\phi(x_1, x_2, x_3, x_4, t)$ of mass $M$ in ${\bf R}^3\times {\bf S}^1$, where ${\bf S}^1$ is compactified into a circle of circumference $b$. Two square parallel plates of area $L^2$ are present in ${\bf R}^3$ and are perpendicular to the $x_3$-axis and located at $x_3=0$ and $x_3=a$. The following periodic boundary condition on $\phi$
\begin{equation}
\phi(x_1, x_2, x_3, 0, t)=\phi(x_1, x_2, x_3, b, t)
\label{bc1}
\end{equation}
implements the presence of the compactified extra-dimension  ${\bf S}^1$ \cite{Elizalde:1988rh}. In addition to the boundary condition of Eq. (\ref{bc1}), I will impose either Dirichlet boundary conditions
\begin{equation}
\phi(x_1, x_2, 0, x_4, t)=\phi(x_1, x_2, a, x_4, t)=0,
\label{dirichlet}
\end{equation}
or mixed boundary conditions
\begin{equation}
\phi(x_1, x_2, 0, x_4, t)={\partial\phi\over\partial x_3}(x_1, x_2, a, x_4, t)=0,
\label{mixed}
\end{equation}
on the scalar field at the plates. I will not investigate Neumann boundary conditions since they produce the same results as Dirichlet ones. My goal is to understand how the presence of one compactified extra dimension modifies the Casimir effect due to the complex scalar field, and will use the generalized zeta function technique to investigate this problem.

I use Euclidean time $\tau$ and begin by writing the Klein Gordon operator
\begin{equation}
D_E=-{\partial^2\over \partial \tau^2}-\nabla^2-{\partial^2\over \partial x_4^2}+M^2,
\label{KGoperator}
\end{equation}
where the contribution of the compactified extra dimension is shown explicitly.  The eigenvalues of $D_E$ are:
\begin{equation}
\left\{k_0^2+k_1^2+k_2^2+k_3^2+\left({2m\pi\over b}\right)^2+M^2\right\},
\label{eigenvalues1}
\end{equation}
where $k_0, k_1, k_2 \in \Re$, $m=0,\pm 1,\pm 2,\cdots$, and $k_3$ takes the following discrete values when considering Dirichlet boundary conditions
\begin{equation}
k_3={n\pi\over a},
\label{k31}
\end{equation}
with $n=1,2,\cdots$. When mixed boundary conditions are considered, $k_3$ takes different discrete values
\begin{equation}
k_3=\left(n+\textstyle{\frac{1}{2}}\right){\pi\over a},
\label{k32}
\end{equation}
with $n=0,1,2,\cdots$.

Using the eigenvalues of Eq. (\ref{eigenvalues1}) and $k_3$ of Eq. (\ref{k31}), I construct the zeta function for the case of Dirichlet boundary conditions 
\begin{equation}
\zeta(s)=\mu^{2s}{L^2\over (2\pi)^3}\int d^3k \sum_{n=1}^\infty \sum_{m=-\infty}^\infty \left[k^2+ \left({n\pi\over a}\right)^2+\left({2m\pi\over b}\right)^2+M^2\right]^{-s},
\label{zetad1}
\end{equation}
where $d^3k=dk_0 dk_1 dk_2$, $k^2=k_0^2+k_1^2+k_2^2$ and, as it is done routinely when using the zeta function technique, I introduce the multiplicative parameter $\mu$ with dimension of mass to keep the zeta function with dimensions of energy for all values of $s$ \cite{Hawking:1976ja}.
For mixed boundary conditions, I use Eqs. (\ref{eigenvalues1}) and (\ref{k32}) and find
\begin{equation}
\zeta(s)=\mu^{2s}{L^2\over (2\pi)^3}\int d^3k \sum_{n=0}^\infty \sum_{m=-\infty}^\infty \left[k^2+ \left(n+\textstyle{\frac{1}{2}}\right)^2\left({\pi\over a}\right)^2+\left({2m\pi\over b}\right)^2+M^2\right]^{-s}.
\label{zetam1}
\end{equation}
\section{Dirichlet boundary conditions}
\label{3}
When investigating Dirichlet boundary conditions, I use the following identity
\begin{equation}
z^{-s}=\frac{1}{ \Gamma(s)}\int_0^\infty dt\, t^{s-1}e^{-zt}
\label{z-s}
\end{equation}
in Eq. (\ref{zetad1}), then do the straightforward $k$-integration to obtain
\begin{equation}
\zeta(s)={\mu^{2s}\over \Gamma(s)}{L^2\over (2\sqrt{\pi})^3}\sum_{n=1}^\infty \sum_{m=-\infty}^\infty\int_0^\infty dt \,\,t^{s-5/2} 
e^{-({n^2\pi^2\over a^2}+{4m^2\pi^2\over b^2}+M^2)t}.
\label{zetad2}
\end{equation}
At this point I do a Poisson resummation of the $n$-sum
\begin{equation}
\sum_{n=1}^\infty e^{-{n^2\pi^2\over a^2}t} = {a\over 2\sqrt{\pi t}}\sum_{n=-\infty}^\infty e^{-{n^2a^2\over t}}-{\frac{1}{2}},
\label{Poisson1}
\end{equation}
drop the $-{1\over 2}$ term since it does not depend on the plate distance $a$, and obtain
\begin{equation}
\zeta(s)={\mu^{2s}\over \Gamma(s)}{L^2a\over (4\pi)^2}\int_0^\infty dt \,\,t^{s-3}e^{-M^2t} \left(1+2\sum_{n=1}^\infty e^{-{n^2a^2\over t}}\right)\left(1+2\sum_{m=1}^\infty e^{-{4m^2\pi^2\over b^2}t}\right).
\label{zetad3}
\end{equation}
Terms with only a linear dependence on $a$ can be ignored, since they are uniform energy density terms that do not depend on the presence of the plates, and I write
\begin{equation}
\zeta(s)=\zeta_1(s)+\zeta_2(s),
\label{zetad4}
\end{equation}
where
\begin{equation}
\zeta_1(s)={\mu^{2s}\over \Gamma(s)}{L^2a\over 8\pi^2}\int_0^\infty dt \,\,t^{s-3}e^{-M^2t} \sum_{n=1}^\infty e^{-{n^2a^2\over t}},
\label{zetad5}
\end{equation}
depends only on the plate distance $a$, and $\zeta_2(s)$ depends also on $b$, the compactification length of the large extra dimension
\begin{equation}
\zeta_2(s)={\mu^{2s}\over \Gamma(s)}{L^2a\over 4\pi^2}\int_0^\infty dt \,\,t^{s-3}e^{-M^2t} \sum_{n=1}^\infty e^{-{n^2a^2\over t}}\sum_{m=1}^\infty e^{-{4m^2\pi^2\over b^2}t}.
\label{zetad6}
\end{equation}
I am going to investigate first the light mass scenario, $M<a^{-1}$, then the heavy mass one, $M>a^{-1}$.

In the case of low mass, I do a power series expansion of the $e^{-M^2t}$ term in $\zeta_1(s)$ and retain the first three terms, change the integration variable, ${n^2a^2\over t} = t'$, to obtain
\begin{equation}
\zeta_1(s)={(\mu a)^{2s}\over \Gamma(s)}{L^2\over 8\pi^2a^3}\left[\zeta_R(4-2s)\Gamma(2-s)-M^2a^2\zeta_R(2-2s)\Gamma(1-s) +{M^4a^4\over 2}\zeta_R(-2s)\Gamma(-s)\right],
\label{zetad7}
\end{equation}
where $\zeta_R(s)$ is the Riemann zeta function and $\Gamma(s)$ is the Euler gamma function. The Casimir energy $E$ is obtained  from the zeta function by taking its derivative at $s=0$
\begin{equation}
E=-\zeta'(0),
\label{EC1}
\end{equation}
therefore it is sufficient to evaluate $\zeta(s)$ when $s$ is near zero. For $s$ close to zero I find
\begin{equation}
{(\mu a)^{2s}\over \Gamma(s)}\zeta_R(4-2s)\Gamma(2-s)\simeq {\pi^4\over 90}s+{\cal O}(s^2),
\label{id1}
\end{equation}
\begin{equation}
{(\mu a)^{2s}\over \Gamma(s)}\zeta_R(2-2s)\Gamma(1-s)\simeq {\pi^2\over 6}s+{\cal O}(s^2),
\label{id2}
\end{equation}
\begin{equation}
{(\mu a)^{2s}\over \Gamma(s)}\zeta_R(-2s)\Gamma(-s)\simeq \left({1\over 2s}+\gamma_E+\log{\mu a\over 2\pi}\right)s+{\cal O}(s^2),
\label{id3}
\end{equation}
where $\gamma_E$ is the Euler-Mascheroni constant, and obtain $E_1=-\zeta'_1(0)$, which is the contribution of $\zeta_1(s)$ to the Casimir energy
\begin{equation}
E_1=-{L^2\over 8\pi^2a^3}\left[{\pi^4\over 90}-{\pi^2\over 6}M^2a^2+{M^4a^4\over2}\log (Ma)
\right]
\label{E1d}
\end{equation}
where I neglected uniform energy density terms that do not contribute to the Casimir energy and made the obvious choice $\mu = M$ indicating on-shell renormalization.
Next, I evaluate $\zeta_2(s)$ by changing integration variable, $t\rightarrow {nab\over 2\pi m}t$, and obtain
\begin{equation}
\zeta_2(s)=\sum_{m,n=1}^\infty\left({\mu^2nab\over 2\pi m}\right)^s{L^2a\over 4\pi^2\Gamma(s)}\left({2\pi m\over n a b}\right)^2\int_0^\infty dt \,\,t^{s-3}e^{-(M^2{nab\over 2\pi m})t}  e^{-{2\pi mn a\over b}(t+t^{-1})}.
\label{zeta2d}
\end{equation}
As long as ${a\over b}>{1\over 2\pi}$, I can evaluate this integral using the saddle point method and, for $s\ll1$, obtain
\begin{equation}
\zeta_2(s)\simeq{L^2\over\sqrt{2}(ab)^{3/2}}s\sum_{m,n=1}^\infty{ m^{3/2}\over n^{5/2}}e^{-(M^2{nab\over 2\pi m})}  e^{-{4\pi mn a\over b}},
\label{zeta2d2}
\end{equation}
whose contribution to the Casimir energy, $E_2=-\zeta'_2(0)$, is
\begin{equation}
E_2=-{L^2\over\sqrt{2}(ab)^{3/2}}\sum_{m,n=1}^\infty{ m^{3/2}\over n^{5/2}}e^{-(M^2{nab\over 2\pi m})}  e^{-{4\pi mn a\over b}}.
\label{E2d}
\end{equation}
Below is the Casimir energy, $E=E_1+E_2$, per unit area, in the case of light mass
\begin{equation}
{E\over L^2}=-{\pi^2\over 720a^3}\left[1-{15\over \pi^2}y^2+{45\over\pi^4}y^4\log y+{720\over\sqrt{2}\pi^2}x^{3/2}\sum_{m,n=1}^\infty{ m^{3/2}\over n^{5/2}}e^{-{ny^2\over 2\pi mx}}  e^{-{4\pi mn x}}\right],
\label{Ed}
\end{equation}
where the overall factor $-{\pi^2\over 720a^3}$ is the three dimensional Casimir energy per unit area due to a massless scalar, and I introduced the two dimensionless parameters $x={a\over b}$ and $y=Ma$. In the case of light mass being examined now, $y<1$. Notice that Eq. (\ref{Ed}) is valid, and exact, for $x>{1\over 2\pi}$. In the case of a small or non existent extra dimension, $x$ will be very large and the last term in the brackets will vanish, yielding the well known result for the Casimir energy of a complex scalar field with light mass. Only this last term in the brackets depends on $x$ and it contains the entire dependence of $E$ on the compactification length $b$ of the extra dimension. To understand how the presence of the large extra dimension affects the Casimir energy, it is convenient to introduce the following dimensionless function
\begin{equation}
F_D(y,x)=1-{15\over \pi^2}y^2+{45\over\pi^4}y^4\log y+{720\over\sqrt{2}\pi^2}x^{3/2}\sum_{m,n=1}^\infty{ m^{3/2}\over n^{5/2}}e^{-{ny^2\over 2\pi mx}}  e^{-{4\pi mn x}},
\label{Fd}
\end{equation}
which allows me to write the Casimir energy per unit area as
\begin{equation}
{E\over L^2}=-{\pi^2\over 720a^3}F_D(y,x).
\label{Ed2}
\end{equation}
Figure 1, below, displays $F_D(y,x)$ as a function of $x$ for three different values of $y$ and shows that a deviation from the standard three-dimensional Casimir energy starts to happen when $a\sim 0.65 b$ for all three values of $y$ and rapidly reaches an increase of more than $10\%$ over its three-dimensional value when the plate distance $a\sim 0.40 b$. 
\begin{figure}[H]
\centerline{\includegraphics[width=12.0cm]{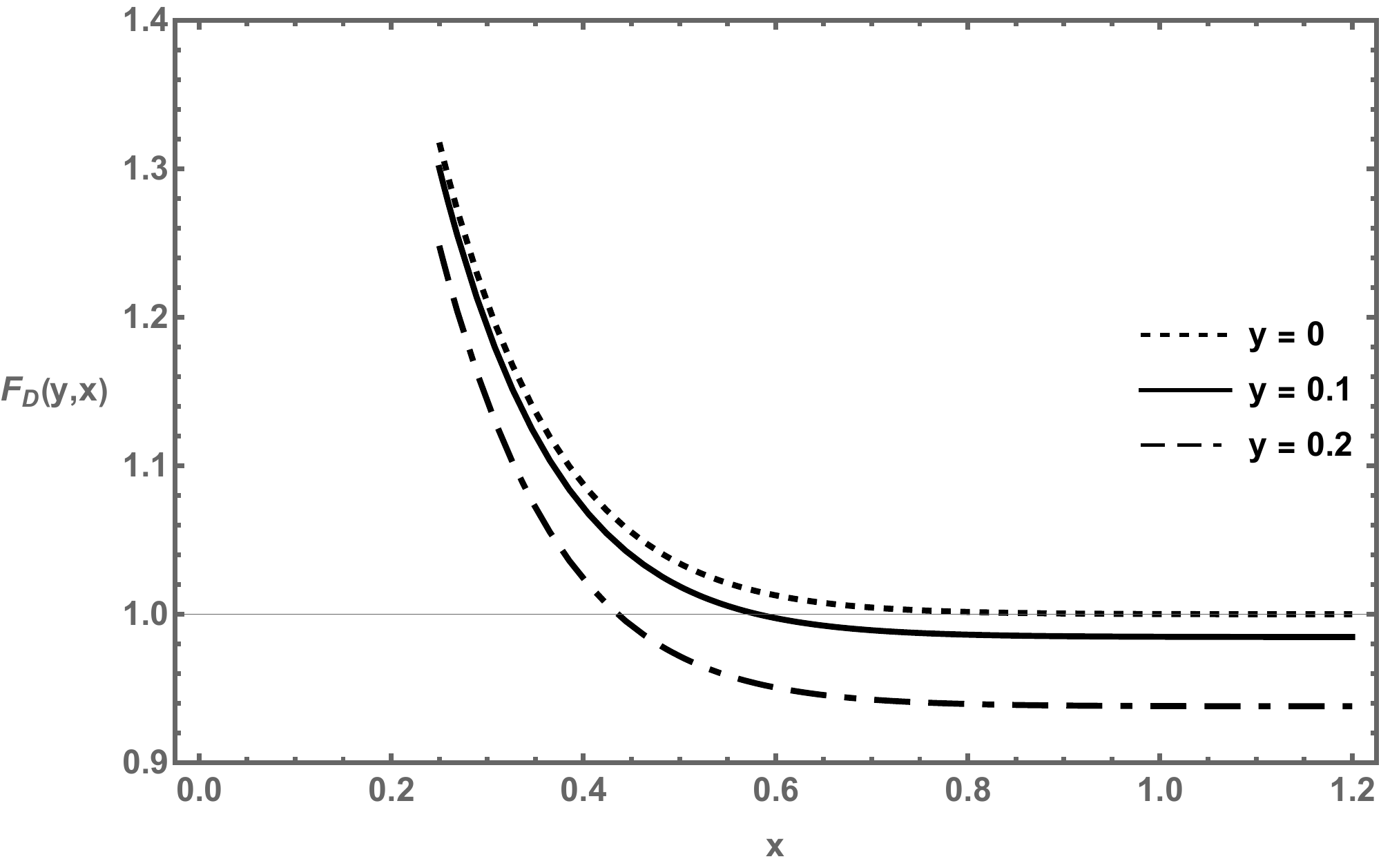}}
\caption{
Plot of $F_D(y,x)$ as a function of $x={a\over b}$, for $y=0$, $y=0.1$, and $y=0.2$.
\label{fig1}
           }
\end{figure}
If a large compactified extra dimension is not present, then $x\rightarrow \infty$ and $F_D(y,\infty)$ represents the correction to the three dimensional Casimir energy due to a non-vanishing, but light, scalar mass. For a massless scalar field, $y=0$ and $F_D(0,\infty)=1$. Figure 2 displays the ratio $F_D(y,x)\over F_D(y,\infty)$ as a function of $x$ for the same three values of $y$ used in above.
\begin{figure}[h]
\centerline{\includegraphics[width=12.0cm]{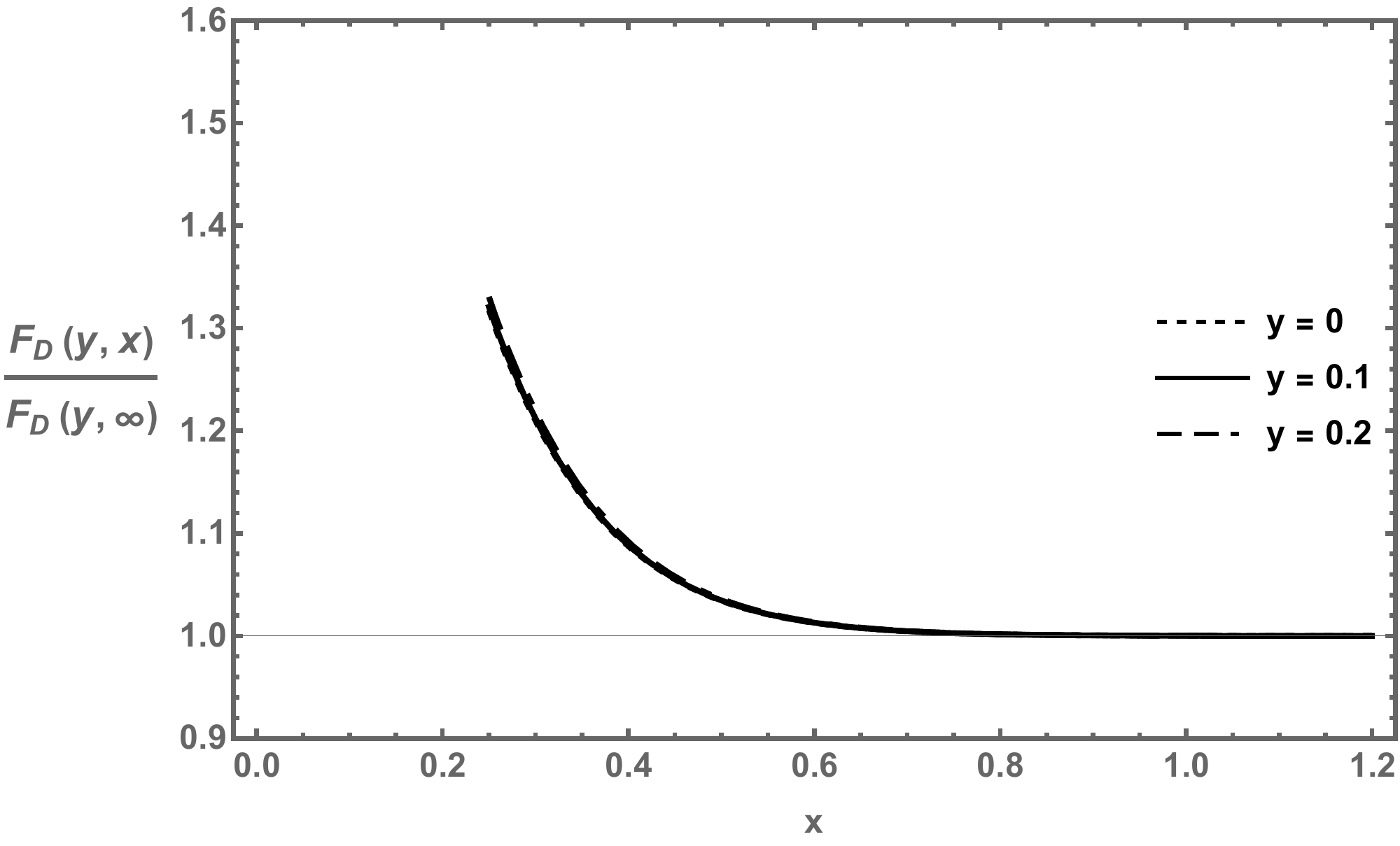}}
\caption{
Plot of $F_D(y,x)\over F_D(y,\infty)$  as a function of $x={a\over b}$, for $y=0$, $y=0.1$, and $y=0.2$.
\label{fig2}
           }
\end{figure}
Notice that the three curves are identical and therefore completely superimposed onto each other, indicating that the effect of the extra dimension on the Casimir energy appears for the same plate distance value regardless of the value of the scalar field mass, as long as it is a light mass. The value of $F_D(y,x)\over F_D(y,\infty)$ is $1.1$ for $x\sim0.4$, indicating that, when $a\sim0.4b$, the Casimir energy in the presence of the extra dimension increases by $10\%$ compared to its value in the absence of the extra dimension. The increase rises to $20\%$ when $x\sim 0.3$.

Next, I investigate the case of large mass, $M>a^{-1}$. I start with the expression of $\zeta_1(s)$ from Eq. (\ref{zetad5}), change variable of integration, $t\rightarrow {na\over M}t$, and find
\begin{equation}
\zeta_1(s)= {L^2\over 8\pi^2\Gamma(s)}\sum_{n=1}^\infty\left({n\mu^{2}a\over M}\right)^s{M^2\over n^2a }\int_0^\infty dt \,\,t^{s-3}e^{-nMa(t+t^{-1})}.
\label{zeta1d1}
\end{equation}
Since $Ma>1$, I integrate using the saddle point method and obtain
\begin{equation}
\zeta_1(s)= {L^2\over 8\pi^{3/2}\Gamma(s)}\left({M\over a }\right)^{3/2}\sum_{n=1}^\infty\left({n\mu^{2}a\over M}\right)^s{e^{-2nMa}\over n^{5/2}}.
\label{zeta1d2}
\end{equation}
I continue and evaluate the expression of $\zeta_2(s)$ from Eq. (\ref{zetad6}). I change variable of integration, $t\rightarrow {na\over \sqrt{M^2+{4\pi^2m^2\over b^2}}}t$, to obtain
\begin{equation}
\zeta_2(s)= {L^2\over 4\pi^2\Gamma(s)}\sum_{m=1}^\infty\sum_{n=1}^\infty\left({n\mu^{2}a\over \sqrt{M^2+{4\pi^2m^2\over b^2}}}\right)^s\left({M^2+{4\pi^2m^2\over b^2}\over n^2a }\right)\int_0^\infty dt \,\,t^{s-3}e^{-na\sqrt{M^2+{4\pi^2m^2\over b^2}}(t+t^{-1})},
\label{zeta2d1}
\end{equation}
then integrate using the saddle point method and find
\begin{equation}
\zeta_2(s)= {L^2\over 4\pi^{3/2}\Gamma(s)}\sum_{m=1}^\infty\sum_{n=1}^\infty\left({n\mu^{2}a\over \sqrt{M^2+{4\pi^2m^2\over b^2}}}\right)^s\left({\sqrt{M^2+{4\pi^2m^2\over b^2}}\over a }\right)^{3/2}{e^{-2na\sqrt{M^2+{4\pi^2m^2\over b^2}}}\over n^{5/2}}.
\label{zeta2d3}
\end{equation}
The Casimir energy $E$ is obtained from $\zeta_1(s)$ and $\zeta_2(s)$ by using $E=-\zeta_1'(0)-\zeta_2'(0)$, and I find
\begin{equation}
{E\over L^2}= -{1\over 8\pi^{3/2}}\left({M\over a }\right)^{3/2}\left[\sum_{n=1}^\infty{e^{-2nMa}\over n^{5/2}}+2\sum_{m=1}^\infty\sum_{n=1}^\infty\left({1+{4\pi^2m^2\over M^2b^2} }\right)^{3/4}{e^{-2na\sqrt{M^2+{4\pi^2m^2\over b^2}}}\over n^{5/2}}\right].
\label{Ed3}
\end{equation}
Since $Ma>1$, only terms with $n=1$ will contribute significantly and, using the two dimensionless parameters $x=a/b$ and $y=Ma$ introduced previously, I obtain
\begin{equation}
{E\over L^2}= -{1\over 8\pi^{3/2}}\left({M\over a }\right)^{3/2}e^{-2Ma}\left[1+2\sum_{m=1}^\infty\left({1+{4\pi^2m^2x^2\over y^2} }\right)^{3/4}e^{-2y\left(\sqrt{1+{4\pi^2m^2x^2\over y^2}}-1\right)}\right].
\label{Ed4}
\end{equation}
Notice that the first of the two terms in Eq. (\ref{Ed4}) is the well known three-dimensional Casimir energy per unit area due to a scalar field with large mass, while the second term is the correction due to the large compactified extra dimension. To facilitate understanding the role of the extra dimension, I introduce the following dimensionless function
\begin{equation}
H(y,x)= 1+2\sum_{m=1}^\infty\left({1+{4\pi^2m^2x^2\over y^2} }\right)^{3/4}e^{-2y\left(\sqrt{1+{4\pi^2m^2x^2\over y^2}}-1\right)},
\label{Hd1}
\end{equation}
which allows me to write the Casimir energy per unit area by factoring out its three dimensional form
\begin{equation}
{E\over L^2}= -{1\over 8\pi^{3/2}}\left({M\over a }\right)^{3/2}e^{-2Ma}H(y,x).
\label{Hd2}
\end{equation}
Figure 3, below, displays $H(y,x)$ as a function of $x$ for three different values of $y$ showing a deviation from the standard three dimensional Casimir effect when $y=1$ for $x\sim 0.7$, when $y=5$ for $x\sim 1$, and when $y=10$ for $x\sim 1.3$.
\begin{figure}[H]
\centerline{\includegraphics[width=12.0cm]{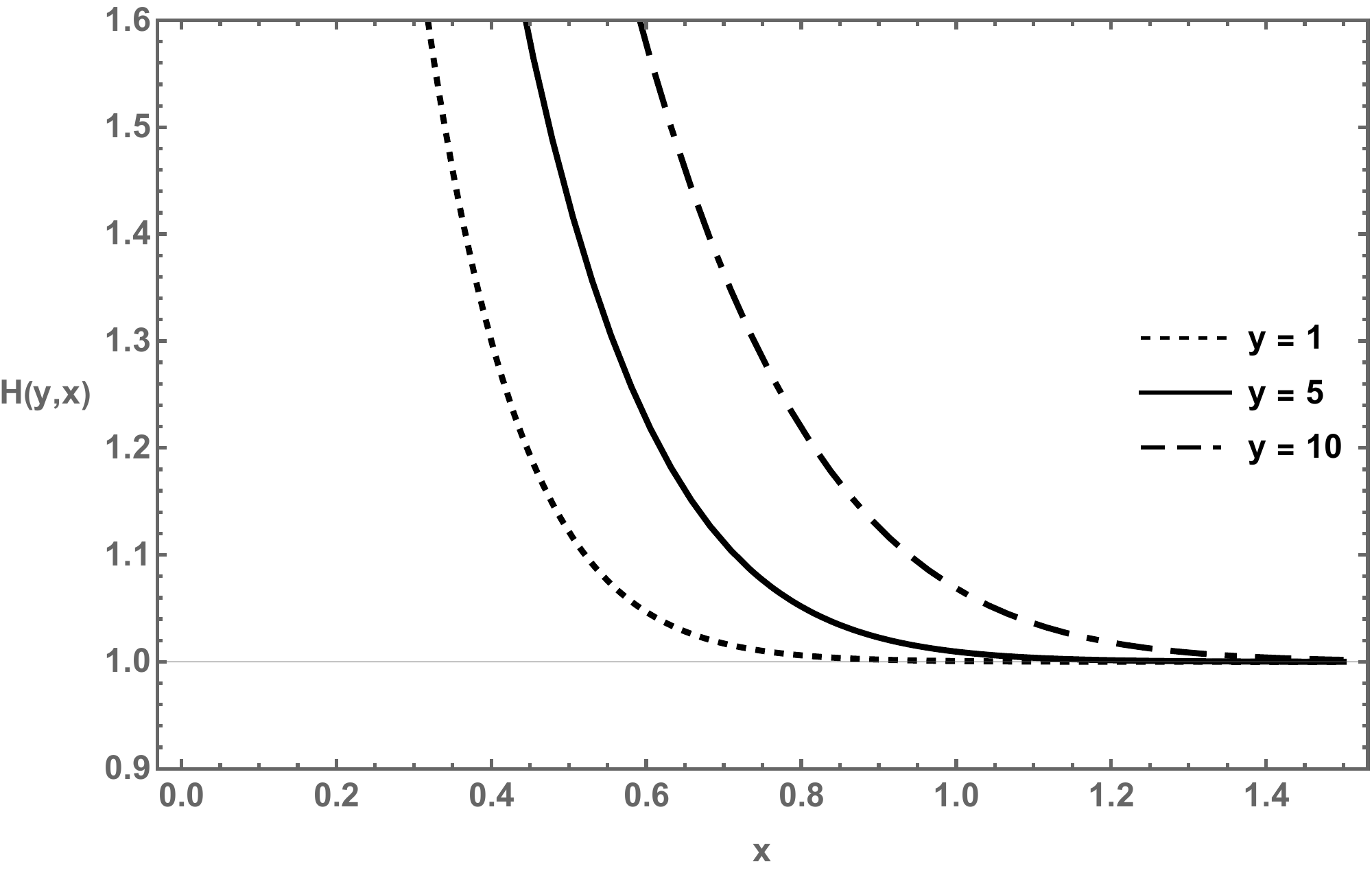}}
\caption{
Plot of $H(y,x)$  as a function of $x={a\over b}$, for $y=1$, $y=5$, and $y=10$.
\label{fig3}
           }
\end{figure}
When comparing Figs. 1 and 3 it is clear that, in the large scalar mass scenario, significant deviations from the standard three dimensional Casimir effect happen for larger values of the plate distance $a$ than they do in the low mass case and, in the large mass case, deviations grow large faster than in the case of light mass.
\section{Mixed boundary conditions}
\label{4}
When investigating mixed boundary conditions, I use identity (\ref{z-s}) in Eq. (\ref{zetam1}), do the $k$-integration and obtain
\begin{equation}
\zeta(s)={\mu^{2s}\over \Gamma(s)}{L^2\over (2\sqrt{\pi})^3}\sum_{n=0}^\infty \sum_{m=-\infty}^\infty\int_0^\infty dt \,\,t^{s-5/2} 
e^{-\left[(n+{1\over 2})^2({\pi\over a})^2+{4m^2\pi^2\over b^2}+M^2\right]t}.
\label{zetam2}
\end{equation}
Next, I do a Poisson resummation of the $n$-sum
\begin{equation}
\sum_{n=0}^\infty e^{-(n+{1\over 2})^2({\pi\over a})^2t} = {a\over 2\sqrt{\pi t}}\sum_{n=-\infty}^\infty (-1)^ne^{-{n^2a^2\over t}},
\label{Poisson2}
\end{equation}
and obtain
\begin{equation}
\zeta(s)={\mu^{2s}\over \Gamma(s)}{L^2a\over (4\pi)^2}\int_0^\infty dt \,\,t^{s-3}e^{-M^2t} \left(1+2\sum_{n=1}^\infty (-1)^ne^{-{n^2a^2\over t}}\right)\left(1+2\sum_{m=1}^\infty e^{-{4m^2\pi^2\over b^2}t}\right).
\label{zetam3}
\end{equation}
Neglecting uniform energy density terms with only a linear dependence on $a$, I write again $\zeta(s)=\zeta_1(s)+\zeta_2(s)$, where $\zeta_1(s)$ depends only on the plate distance $a$
\begin{equation}
\zeta_1(s)={\mu^{2s}\over \Gamma(s)}{L^2a\over 8\pi^2}\int_0^\infty dt \,\,t^{s-3}e^{-M^2t} \sum_{n=1}^\infty (-1)^ne^{-{n^2a^2\over t}},
\label{zetam5}
\end{equation}
and $\zeta_2(s)$ depends also on the compactification length $b$ of the large extra dimension
\begin{equation}
\zeta_2(s)={\mu^{2s}\over \Gamma(s)}{L^2a\over 4\pi^2}\int_0^\infty dt \,\,t^{s-3}e^{-M^2t} \sum_{n=1}^\infty (-1)^ne^{-{n^2a^2\over t}}\sum_{m=1}^\infty e^{-{4m^2\pi^2\over b^2}t}.
\label{zetam6}
\end{equation}
In the low mass scenario, I do a power series expansion of $e^{-M^2t}$ in $\zeta_1(s)$ and retain the first three terms, change the integration variable as in the previous section, and obtain
\begin{eqnarray}
\zeta_1(s)&=&{(\mu a)^{2s}\over \Gamma(s)}{L^2\over 8\pi^2a^3}\biggl[(2^{2s-3}-1)\zeta_R(4-2s)\Gamma(2-s)-M^2a^2(2^{2s-1}-1)\zeta_R(2-2s)\Gamma(1-s)
\nonumber \\
&+&\left.
{M^4a^4\over 2}(2^{2s+1}-1)\zeta_R(-2s)\Gamma(-s)\right],
\label{zetam7}
\end{eqnarray}
which, using the following identities valid for $s\ll 1$
\begin{equation}
{(\mu a)^{2s}\over \Gamma(s)}(2^{2s-3}-1)\zeta_R(4-2s)\Gamma(2-s)\simeq -{7\pi^4\over 720}s+{\cal O}(s^2),
\label{id4}
\end{equation}
\begin{equation}
{(\mu a)^{2s}\over \Gamma(s)}(2^{2s-1}-1)\zeta_R(2-2s)\Gamma(1-s)\simeq -{\pi^2\over 12}s+{\cal O}(s^2),
\label{id5}
\end{equation}
\begin{equation}
{(\mu a)^{2s}\over \Gamma(s)}(2^{2s+1}-1)\zeta_R(-2s)\Gamma(-s)\simeq \left({1\over 2s}+\gamma_E+\log{2\mu a\over \pi}\right)s+{\cal O}(s^2),
\label{id6}
\end{equation}
contributes an amount $E_1$, shown below, to the Casimir energy
\begin{equation}
E_1={7\pi^2\over 5,760 }{L^2\over a^3}\left[1-{60\over 7\pi^2}(Ma)^2-{360\over 7\pi^4}(Ma)^4\log (Ma)\right],
\label{E1m1}
\end{equation}
where a uniform energy density term has been neglected. The second part of the zeta function, $\zeta_2(s)$, is obtained as it is done in the previous section, with an extra factor of $(-1)^n$ being the only difference, and its contribution to the Casimir energy is
\begin{equation}
E_2=-{L^2\over\sqrt{2}(ab)^{3/2}}\sum_{m,n=1}^\infty(-1)^n{ m^{3/2}\over n^{5/2}}e^{-(M^2{nab\over 2\pi m})}  e^{-{4\pi mn a\over b}}.
\label{E2m}
\end{equation}
The Casimir energy in the case of low mass, $E=E_1+E_2$, can be written as
\begin{equation}
{E\over L^2}={7\pi^2\over 5,760a^3}F_m(y,x),
\label{Em2}
\end{equation}
where
\begin{equation}
F_m(y,x)=1-{60\over 7\pi^2}y^2-{360\over 7\pi^4}y^4\log y-{5,760\over7\sqrt{2}\pi^2}x^{3/2}\sum_{m,n=1}^\infty (-1)^n{ m^{3/2}\over n^{5/2}}e^{-{ny^2\over 2\pi mx}}  e^{-{4\pi mn x}}.
\label{Fm}
\end{equation}
Figure 4 displays $F_m(y,x)$ as a function of $x$ for the same three values of $y$ used previously and shows, one more time, that a deviation from the standard mixed boundary conditions Casimir effect starts to happen when $x\sim 0.65$ for all three values of $y$. Notice that an increase of about $10\%$ of the Casimir energy over its three-dimensional value is observed when $x\sim 0.4$, indicating again a rapid deviation of the Casimir effect with one compactified extra dimension from the standard effect also for mixed boundary conditions. Finally, it is important to point out that the repulsive nature of the Casimir effect with mixed boundary conditions is maintained and, for small $x$, the repulsion is enhanced. If this feature of the repulsive Casimir effect is confirmed also for the spherical geometry, it could have interesting cosmological implications. An investigation of the spherical Casimir effect in the presence of a large compactified extra dimension is forthcoming.

As it happens in the case of Dirichlet boundary conditions, the three curves of the ratio $F_m(y,x)\over F_m(y,\infty)$ as a function of $x$ are identical to each other and to the curve  $F_m(0,x)$ shown in Figure 4, indicating that the Casimir energy in the presence of the compactified extra dimension exceeds the three-dimensional Casimir energy by $10\%$ or more when $a\sim 0.4 b$ or less.
\begin{figure}[H]
\centerline{\includegraphics[width=12.0cm]{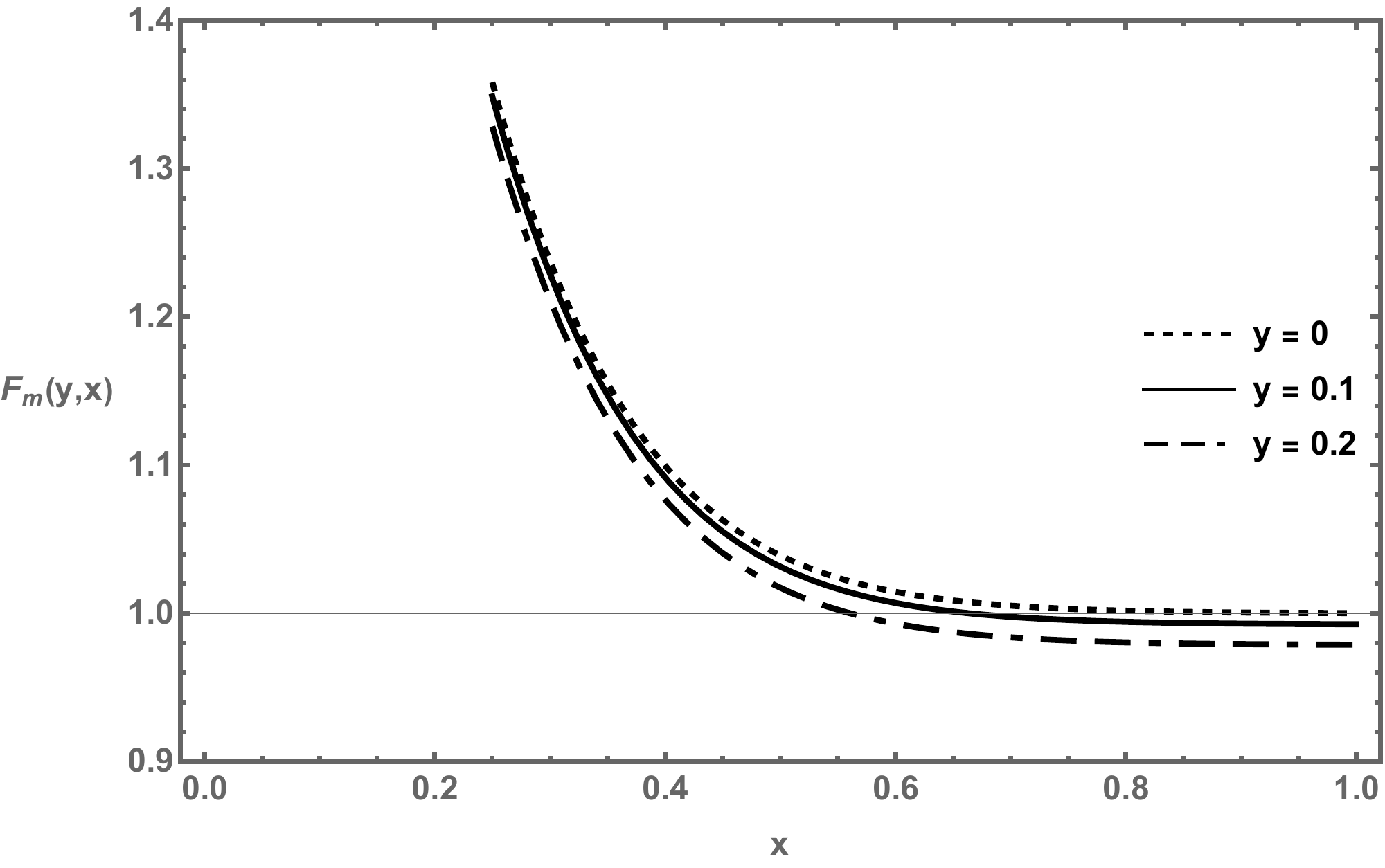}}
\caption{
Plot of $F_m(y,x)$ as a function of $x={a\over b}$, for $y=0$, $y=0.1$, and $y=0.2$.
\label{fig4}
           }
\end{figure}

The large mass limit, $M>a^{-1}$, for mixed boundary conditions is obtained by changing integration variable in $\zeta_1(s)$, $t\rightarrow {na\over M}t$, and in $\zeta_2(s)$, $t\rightarrow {na\over \sqrt{M^2+{4\pi^2m^2\over b^2}}}t$. Following the same steps taken in the case of Dirichlet boundary conditions, I find
\begin{equation}
{E\over L^2}= {1\over 8\pi^{3/2}}\left({M\over a }\right)^{3/2}e^{-2Ma}H(y,x),
\label{Hm}
\end{equation}
where the only difference with the Dirichlet boundary conditions case is the lack of the overall negative sign, as expected for mixed boundary conditions. The function $H(y,x)$ is displayed in Fig. 3 for three different values of $y$. Notice that the overall positive sign for the case of mixed boundary conditions produces a repulsive Casimir force, as it is well known. Notice also that, for $x\rightarrow \infty$, the well known result for mixed boundary conditions in the three-dimensional case is recovered.
\section{Casimir pressure}
\label{5}
The Casimir pressure is obtained by taking a derivative of the Casimir energy
\begin{equation}
P=-{1\over L^2}{\partial E\over\partial a}.
\label{P_1}
\end{equation}
For Dirichlet boundary conditions and light mass, I find
\begin{equation}
P=-{\pi^2\over 240 a^4}G_D(y,x),
\label{P1}
\end{equation}
with
\begin{equation}
G_D(y,x)=1-{5\over \pi^2}y^2-{15\over\pi^4}y^4(1+\log y)+{240\over\sqrt{2}\pi^2}x^{3/2}\sum_{m,n=1}^\infty\left({ m\over n}\right)^{3/2}e^{-{ny^2\over 2\pi mx}}  e^{-{4\pi mn x}}
\left({3\over 2n}+4\pi mx +{ny\over 2\pi m x}\right).
\label{Gd}
\end{equation}
Figure 5 shows $G_D(y,x)$ as a function of $x$ for the same three values of $y$ used previously.  As it is the case for the Casimir energy, we observe a deviation from the standard three-dimensional Casimir pressure starting to happen when $a\sim 0.65 b$ for all three values of $y$ and reaching an increase of more than $10\%$ over its three-dimensional value when the plate distance $a\sim 0.40 b$. 
\begin{figure}[H]
\centerline{\includegraphics[width=12.0cm]{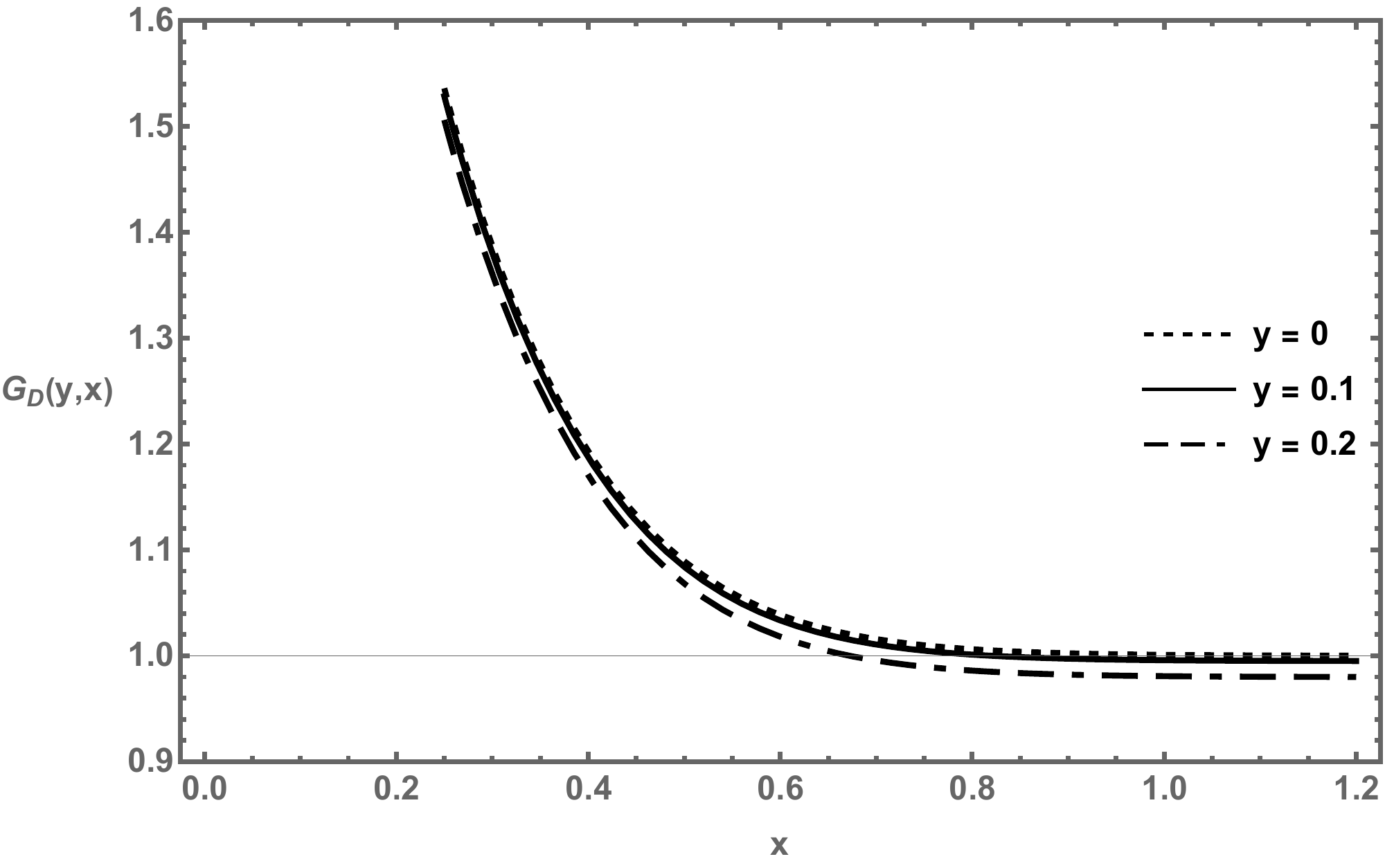}}
\caption{
Plot of $G_D(y,x)$ as a function of $x={a\over b}$, for $y=0$, $y=0.1$, and $y=0.2$.
\label{fig5}
           }
\end{figure}

In the case of Dirichlet boundary conditions and heavy mass, $M>a^{-1}$, I obtain the following 
\begin{equation}
P= -{M^{\frac{5}{2}}\over 4\pi^{\frac{3}{2}}}{e^{-2Ma}\over a^{\frac{3}{2}} }J(y,x),
\label{P2}
\end{equation}
with
\begin{equation}
J(y,x)= 1+2\sum_{m=1}^\infty\left({1+{4\pi^2m^2x^2\over y^2} }\right)^{\frac{5}{4}}e^{-2y\left(\sqrt{1+{4\pi^2m^2x^2\over y^2}}-1\right)}.
\label{J1}
\end{equation}
Figure 6 shows $J(y,x)$ as a function of $x$ for the same three values of $y$ used in the previous section for the case of heavy mass.  A deviation from the standard three-dimensional Casimir pressure starts to happen for $x\sim 0.9$ when $y=1$, for $x\sim 1$ when $y=5$, and for $x\sim 1.3$ when $y=10$. Notice that when $a\gg b$ and thus $x$ is large, the well known Casimir pressure for the three-dimensional case of heavy mass is recovered.
\begin{figure}[H]
\centerline{\includegraphics[width=12.0cm]{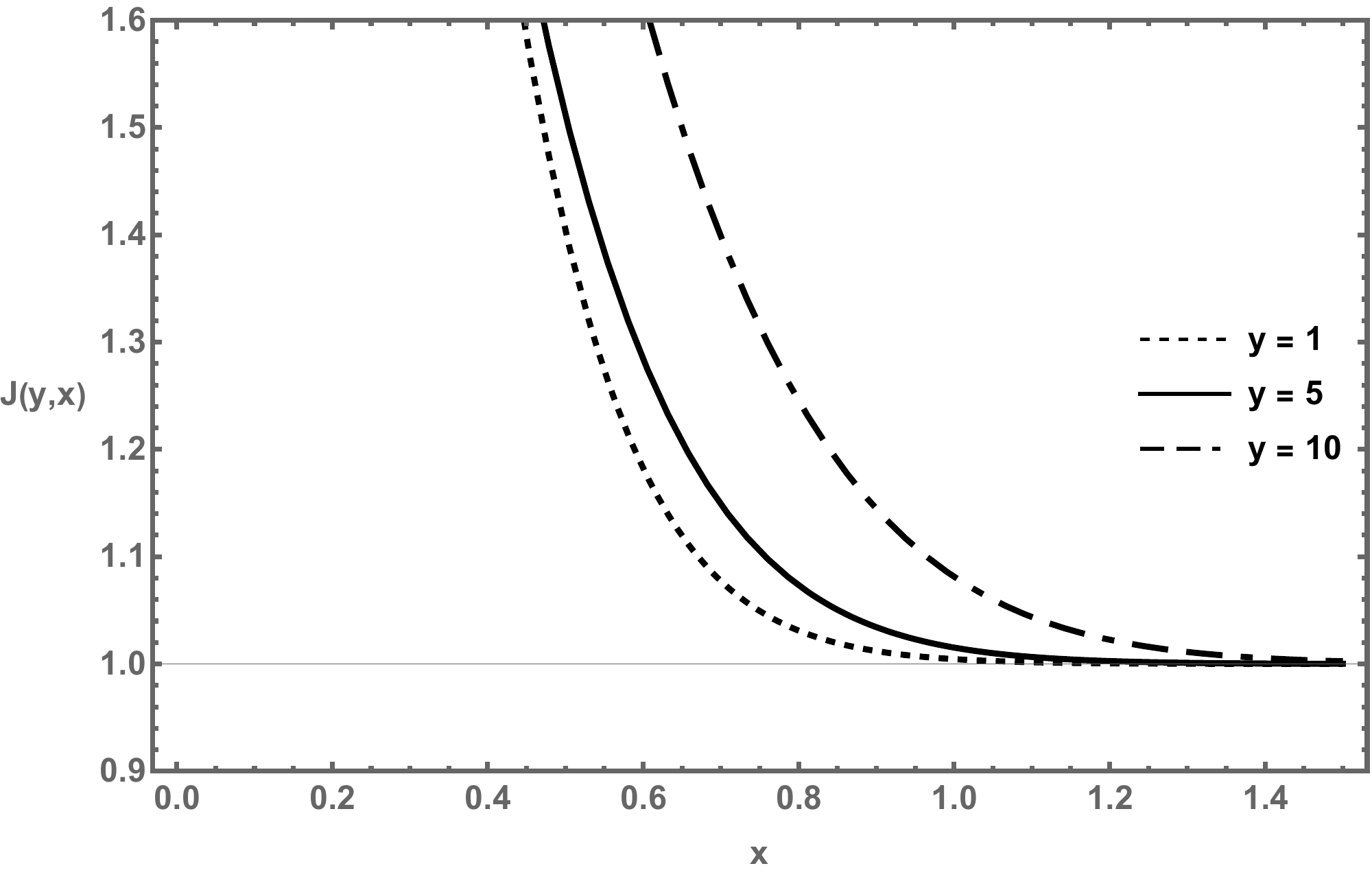}}
\caption{
Plot of $J(y,x)$ as a function of $x={a\over b}$, for $y=1$, $y=5$, and $y=10$.
\label{fig6}
           }
\end{figure}

When investigating mixed boundary conditions and light mass, I obtain
\begin{equation}
P={7\pi^2\over 1,920 a^4}G_m(y,x),
\label{P3}
\end{equation}
where
\begin{equation}
G_m(y,x)=1-{20\over 7\pi^2}y^2+{120\over 7\pi^4}y^4(1+\log y)-{1,920\over 7\sqrt{2}\pi^2}x^{3/2}\sum_{m,n=1}^\infty(-1)^n \left({ m\over n}\right)^{3/2}e^{-{ny^2\over 2\pi mx}}  e^{-{4\pi mn x}}
\left({3\over 2n}+4\pi mx +{ny\over 2\pi m x}\right).
\label{Gm}
\end{equation}
I display $G_m(y,x)$ as a function of $x$ in Figure 7, for the same three values of $y$ used before. Again, we observe a deviation from the standard three-dimensional Casimir pressure starting to happen when $a\sim 0.65b$ for all three values of $y$. It happens also in this case that the well known three dimensional Casimir pressure is obtained when $a\gg b$.
\begin{figure}[H]
\centerline{\includegraphics[width=12.0cm]{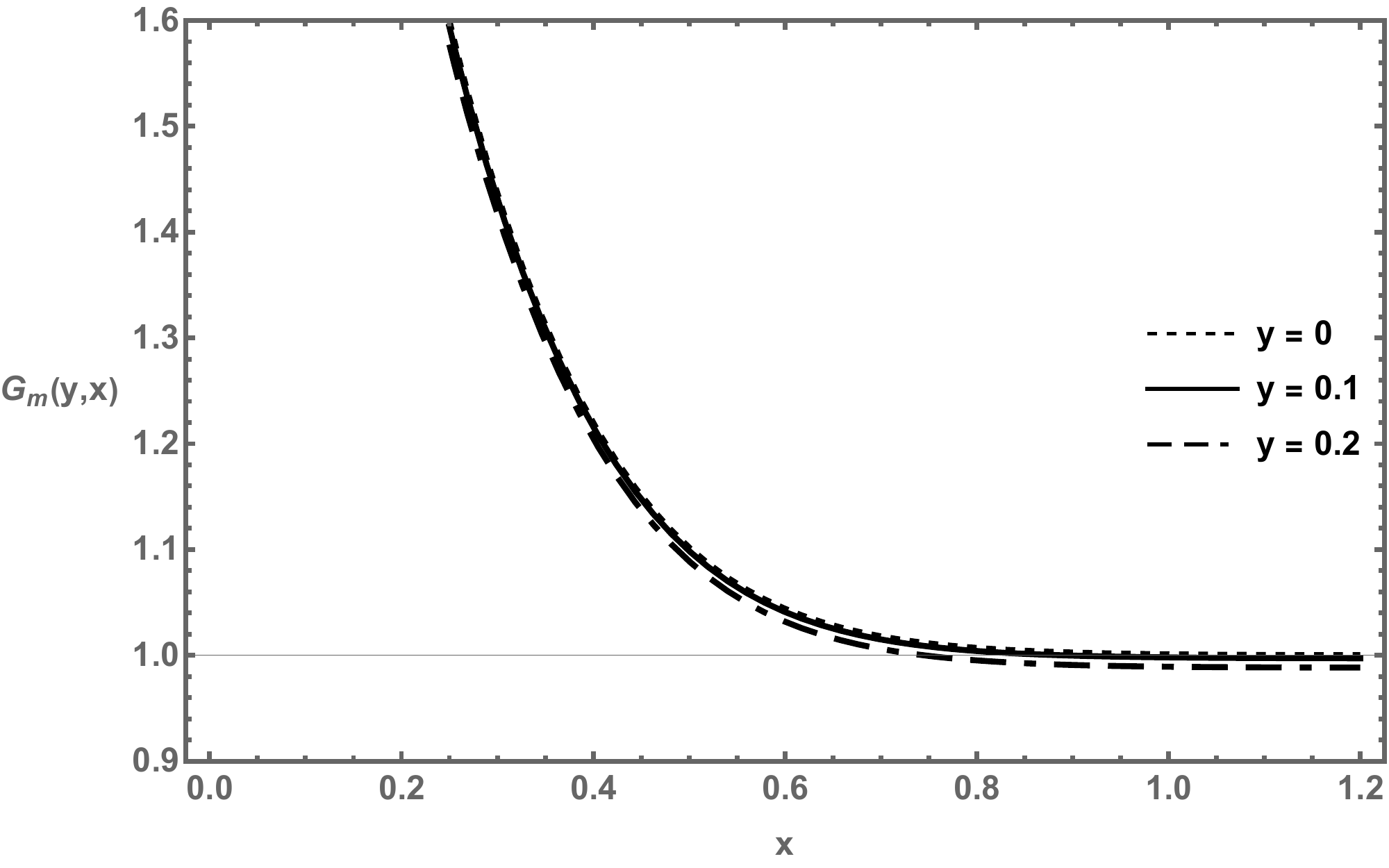}}
\caption{
Plot of $G_m(y,x)$ as a function of $x={a\over b}$, for $y=0$, $y=0.1$, and $y=0.2$.
\label{fig7}
           }
\end{figure}

For mixed boundary conditions and heavy mass, $M>a^{-1}$, I obtain the following 
\begin{equation}
P= {M^{\frac{5}{2}}\over 4\pi^{\frac{3}{2}}}{e^{-2Ma}\over a^{\frac{3}{2}} }J(y,x),
\label{P4}
\end{equation}
where the pressure has the same magnitude as in the case of Dirichlet boundary conditions and heavy mass, but opposite sign. $J(y,x)$ is the same function shown in Eq. (\ref{J1}) for Dirichlet boundary conditions and is displayed in Figure 6.
\section{Discussion and conclusions}
\label{6}

In this work I used the zeta function technique to study the Casimir effect due to a massive complex scalar field that also permeates a compactified fourth dimension. I investigated in detail Dirichlet and mixed boundary conditions. Neumann boundary conditions have not been considered since they produce the same results as Dirichlet.

I obtained simple and exact analytic expressions for the Casimir energy in the case of Dirichlet boundary conditions and low mass, Eq. (\ref{Ed2}) with the auxiliary Eq. (\ref{Fd}) for $F_D(y,x)$, and heavy mass, Eq. (\ref{Hd2}) with the auxiliary Eq. (\ref{Hd1}) for $H_D(y,x)$. Notice that, while the dimensionless functions $F_D$ and $H_D$ contain each an infinite series, these series are quickly convergent and the first two or three terms alone approximate the full sum to within one percent or less. Figs. \ref{fig1} and \ref{fig2} display $F_D(y,x)$ as a function of $x$ for three values of $y$ and show that in the case of low mass, a deviation from the standard Casimir energy starts to happen when $a\sim 0.65 b$, independently of the value of $M$, and quickly increases. Fig. \ref{fig3} displays $H_D(y,x)$ and shows that in the heavy mass scenario, deviations from the standard Casimir energy start to happen when $a\sim 0.9 b$ for $Ma=5$ and when $a\sim 1.2 b$ for $Ma=10$ and increase rapidly. It is clear from the results of this work that the Casimir effect due to a field with heavy mass is affected more strongly by the presence of an accessible compactified fourth dimension. My results for the Casimir pressure, the one quantity measurable in the lab, are also exact simple analytic expressions. In the case of Dirichlet boundary conditions these results are shown in Eqs. (\ref{P1}), (\ref{Gd}), (\ref{P2}), (\ref{J1}), and Figs. \ref{fig5}, \ref{fig6}, confirming fully the onset of a discrepancy with the standard Casimir pressure at the values of $a$ that I report above, and confirming that the Casimir pressure due to a heavily massive scalar field is affected more strongly by the compactified fourth dimension. The infinite series contained in the two dimensionless functions, $G_D(y,x)$ and $J(y,x)$, that appear in my Casimir pressure results are also quickly convergent and can accurately be approximated by their first two or three terms.

I draw the same conclusions from my investigation of the Casimir energy and pressure for the case of mixed boundary conditions. My results show that the Casimir force is repulsive for a scalar field, massive or not, that permeates a compactified extra dimension. Therefore the presence of a compactified extra dimension does not alter the repulsive nature of the mixed boundary conditions effect. The onset of  Casimir energy and pressure deviations from their three-dimensional values happens at the same distances $a$ reported in the previous paragraph, both in the case of light mass and heavy mass, as can be seen from Figs. \ref{fig4} and \ref{fig7} for $F_m$ and $G_m$ respectively.

\end{document}